\documentclass[aps,prl,reprint]{revtex4-1}

\usepackage{graphicx}
\usepackage{amsmath}
\usepackage{sidecap}
\newcommand{\beq}{\begin{eqnarray}}
\newcommand{\eeq}{\end{eqnarray}}

\newcommand{\vp}{{v_\perp}}
\newcommand{\vv}{{\bf v}}

\newcommand{\Phit}{\Phi_{\parallel}}

\newcommand{\scr}[1]{_{\mbox{\protect\scriptsize #1}} }

\newcommand{\B}{{\bf B}}

\def\hblabel#1{\label{#1}}
\newcommand{\x}{{\bf x}}
\newcommand{\vz}{v_{\parallel}}
\newcommand{\fd}[2]{\frac{\displaystyle #1}{\displaystyle #2}}
\newcommand{\cE}{{\cal E}}

\def\eq#1{(\ref{eq:#1})}

\begin{document}

% Use the \preprint command to place your local institutional report
% number in the upper righthand corner of the title page in preprint mode.
% Multiple \preprint commands are allowed.
% Use the 'preprintnumbers' class option to override journal defaults
% to display numbers if necessary
%\preprint{}

%Title of paper
%\title{Spacecraft observations of a Maxwell Demon coating  the separatrix  of asymmetric magnetic reconnection with crescent-shaped  electron distributions}
\title{Spacecraft observations and analytic theory of crescent-shaped electron distributions in asymmetric magnetic reconnection}
% repeat the \author .. \affiliation  etc. as needed
% \email, \thanks, \homepage, \altaffiliation all apply to the current
% author. Explanatory text should go in the []'s, actual e-mail
% address or url should go in the {}'s for \email and \homepage.
% Please use the appropriate macro foreach each type of information

% \affiliation command applies to all authors since the last
% \affiliation command. The \affiliation command should follow the
% other information
% \affiliation can be followed by \email, \homepage, \thanks as well.
%\email[]{Your e-mail address}
%\homepage[]{Your web page}
%\thanks{}
%\altaffiliation{}
\author{J. Egedal$^1$}
\author{A. Le$^2$, W. Daughton$^2$}
\author{B. Wetherton$^1$}
\author{P.~A.  Cassak$^3$}
\author{L.-J Chen$^4$}
\author{B. Lavraud$^5$}
\author{R.~B. Trobert$^6$}
\author{J. Dorelli$^4$}
\author{D.~J. Gershman$^4$}
\author{L.~A. Avanov$^4$}

\affiliation{$^1$Department of Physics, University of Wisconsin-Madison, Madison, Wisconsin, USA}
\affiliation{$^2$Los Alamos National Laboratory, Los Alamos, New Mexico, USA}
\affiliation{$^3$Department of Physics and Astronomy, West Virginia Univ., Morgantown, WV 26506 USA.}
\affiliation{$^4$Heliophysics Science Division, NASA Goddard Space Flight Center, Greenbelt, Maryland, USA}
\affiliation{$^5$Institut de Recherche en Astrophysique et Plan\'etologie, Universit\'e de Toulouse, Toulouse, France}
\affiliation{$^6$University of New Hampshire, Durham, NH, USA}

%\author{W. Daughton}
%\affiliation{Los Alamos National Laboratory, Los Alamos, New Mexico
%87545, USA}
%Collaboration name if desired (requires use of superscriptaddress
%option in \documentclass). \noaffiliation is required (may also be
%used with the \author command).
%\collaboration can be followed by \email, \homepage, \thanks as well.
%\collaboration{}
%\noaffiliation

\date{\today}

\begin{abstract}

Supported by a kinetic simulation, we derive an exclusion energy parameter $\cE_X$ providing a lower kinetic energy bound  for an electron  to cross  from one inflow region to the other during magnetic reconnection. As by a Maxwell Demon, only high energy electrons are permitted to cross the inner reconnection region, setting the electron distribution function observed along the low density side separatrix during asymmetric reconnection. The analytic  model accounts for the  two distinct flavors of crescent-shaped electron distributions observed by spacecraft in a thin boundary layer along the low density separatrix.
\end{abstract}

% insert suggested PACS numbers in braces on next line
% insert suggested keywords - APS authors don't need to do this
%\keywords{}

%\maketitle must follow title, authors, abstract, \pacs, and \keywords
\maketitle

Magnetic reconnection converts magnetic energy into kinetic energy of ions and electrons both during solar flare events \cite{krucker:2010} and reconnection observed {\sl in situ} in Earth's magnetosphere    \cite{oieroset:2001}.
Common for most theoretical models of reconnection is an emphasis on the dynamics of the electrons and their role in breaking the {\sl frozen in} conditions for the electron fluid, permitting the magnetic field lines to change topology and release the stored  magnetic stress in naturally occurring plasmas \cite{dungey:1953}. NASA's new Magnetospheric Multiscale (MMS) mission is specially designed
to address this question, as it can detect {\sl in situ} possible
mechanisms including  electron inertia, pressure tensor effects, and anomalous dissipation for decoupling the electron motion from the magnetic field lines \cite{burch:2016b}.

The identification of diffusion regions in the vast dataset now being recorded by MMS relies in part on numerical and theoretical models for distinct signatures of the reconnection region and the associated separatrix structure.  Recent simulations  of crescent-shaped electron distributions \cite{hesse:2014,chen:2016} have been proposed as a robust signature to find diffusion regions. The crescents are observed in two flavors: perpendicular and parallel to the magnetic field \cite{burch:2016}. The perpendicular crescent shapes  are predicted theoretically in Refs.~\cite{bessho:2016,shay:2016}, using 1D reasoning valid  near the X-line, with electrons interacting strongly with a normal electric field $E_N$.

Considering 2D geometries, here we provide a general derivation which accounts for the occurrence of  both the perpendicular and parallel crescents.
Only high-density (magnetosheath) particles with sufficient energy can cross the diffusion region to the low-density (magnetospheric) inflow region.  Thus, the diffusion region acts like a Maxwell Demon allowing only the most energetic particles across.  This provides an explanation for why the distributions are crescent shaped rather than filled in at lower energies.  The requirement of having a sufficient energy is here quantified in terms of what we call the exclusion energy ${\cE}_X$. As such, magnetosheath electrons with kinetic energies $\cE>\cE_X$ can access magnetic field lines on the magnetospheric side of the separatrix, exiting the region along the separatrix with nearly perfectly  circular perpendicular motion. The parallel streaming and the absence of electrons with $\cE< \cE_X$ yields the parallel crescent-shaped distributions, and their origin is thus different from that proposed in Ref.~\cite{burch:2016}. Contrary to the models applicable near the X-line \cite{bessho:2016,shay:2016}, we find that the perpendicular crescents along the separatrix are comprised of well magnetized electrons with nearly circular perpendicular orbit motion.

\begin{SCfigure*}
%\begin{figure}[h]
\centering
  \includegraphics[width=13 cm]{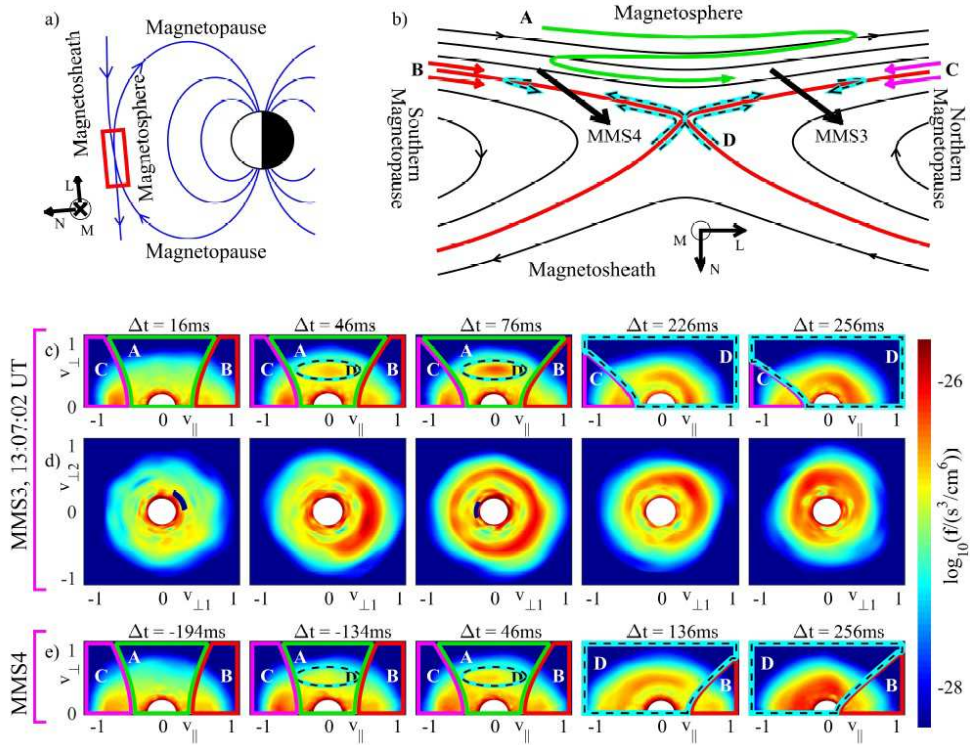}%
  %\vspace{-3.4cm}
\caption{(a) Schematic illustration of the reconnection region encountered by the MMS mission on October 16, 2015.  (b) The trajectories of the MMS spacecraft were determined in \cite{burch:2016} and are indicated by the black arrows. In addition, distinct orbit types are labeled. (c - e) Electron distribution function recorded during the separatrix crossing by MMS3 and MMS4, respectively. Times are given relative to 13:07:02.000 UT. The velocity axes are normalized by $v_0=10^7$m/s, and all distributions are computed from the full 3D MMS measurements. In (c,e), $\bar{f}(\vz,\vp)$  are gyro averaged distributions and regions of distinct orbit types are labeled consistently with the trajectories in (b). The data in  (d)  are cuts of the full 3D electron distributions at $\vz=0$.
}
\label{fig:MMS}
%\end{figure}
\end{SCfigure*}

On October 16, 2015, NASA's MMS mission  had an encounter with an active reconnection region at the dayside magnetopause. The location of the encounter is sketched by the red rectangle in Fig.~\ref{fig:MMS}(a), as was established by the analysis in Ref.~\cite{burch:2016}. Based in part on the recorded time series of the magnetic fields and ion flows, it was concluded  that three of the four MMS spacecraft (MMS1, MMS2, and MMS3) passed the diffusion region on its northern side, while MMS4 passed it on the southern side.  The four spacecraft all recorded similar structures, and here we consider data obtained by MMS3 and MMS4. The paths near the separatrix of these two spacecraft are sketched in Fig.~\ref{fig:MMS}(b), crossing from the low plasma density magnetosphere into the reconnection exhausts, in which the plasma is mainly provided by the much higher densities of the magnetosheath \cite{cassak:2007}.

The distinct types of electron trajectories indicated in Fig.~\ref{fig:MMS}(b) are important to the structures in the electron distribution function. Passing electrons, labeled ${\bf B}$  and  ${\bf C}$, stream into the reconnection region along magnetic field lines, and do not change the signs of their magnetic field aligned (parallel) velocity as they pass through the region. %Thus, these electrons move like beads on strings, with the strings being the instantaneous magnetic field lines.
Trapped electrons, illustrated by the green line labeled ${\bf A}$, have  their  bounce motion caused by trapping by the magnetic mirror force $-\mu\nabla_{\parallel}B$ \cite{lavraud:2016} and the acceleration potential $\Phit=\int_x^{\infty} E_{\parallel} dl$ of Ref.~\cite{egedal:2009pop}. During the course of several bounce motions they convect slowly with the magnetic field lines towards the reconnection separatrix. The basis of this kinetic electron behavior is analogous to that of symmetric reconnection \cite{egedal:jan2005,egedal:2008jgr,egedal:2013}, but for asymmetric reconnection  trapping is most significant in the low density magnetospheric inflow  \cite{egedal:2011popAsym}.
Trajectories of magnetosheath electrons near the separatrix (including their possible reflection back toward the X-line) are schematically illustrated by the cyan-black dashed lines labeled ${\bf D}$ in Fig.~\ref{fig:MMS}(b). %To account for the details in the MMS observations, below we determine how energetic electrons from the magnetosheath can penetrated into the magnetospheric exhaust important to explain the detailed of the MMS electron distributions recorded as the separatrix were crossed.

The distribution functions displayed in Figs.~\ref{fig:MMS}(c-e) are calculated based on the full 3D electron data recorded by the Fast Plasma Investigation (FPI; \cite{pollock:2016}) onboard MMS3 and MMS4 for selected time points  $\Delta t$  relative to 13:07:02.000 UT. In the following we denote gyro-averaged distributions by $\bar{f}$. The distributions $\bar{f}(\vz,\vp)$ in Figs.~\ref{fig:MMS}(c,e) are obtained by first rotating the ``raw'' 3D electron distributions into a coordinate system $(\vz,v_{\perp1},v_{\perp2})$ aligned with the direction of the measured magnetic field. With $\vp=\sqrt{v_{\perp1}^2+v_{\perp2}^2}$, values of $\bar{f}(\vz,\vp)$ are then computed as the average of $f(\vz,v_{\perp1},v_{\perp2})$ over the azimuthal gyroangle $\phi=\tan^{-1}(v_{\perp2}/v_{\perp1})$.

We first consider $\bar{f}(\vz,\vp)$ of MMS3 in Fig.~\ref{fig:MMS}(c) for $\Delta t=16$ ms. Corresponding to the orbit classification in Fig.~\ref{fig:MMS}(b), the regions of trapped electrons are labeled ${\bf A}$, while the regions of passing electrons are labeled by ${\bf B}$ and ${\bf C}$. The trapped passing boundaries are obtained based on the local magnetic field and by estimating $\Phit\simeq45$ V by methods given in \cite{egedal:2010jgr}.
In agreement with Refs.~\cite{egedal:2011popAsym,graham:2016}, this is evidence that the strong parallel electron heating noted in Fig.~3(i) of Ref.~\cite{burch:2016} is mainly due to energization by $\Phit$.

The distributions in Fig.~\ref{fig:MMS}(c) for $\Delta t=46, 76$ ms are similar to that at $\Delta t=16$ ms, except that these include an additional feature within the trapped region. We  mark this feature ${\bf D}$, as  it is caused by energetic magnetosheath electrons penetrating across the separatrix to this location in the magnetospheric inflow. % These electrons thus have the same origin as those labeled ${\bf D}$ at times $\Delta t= 226$ and 256 ms in Fig.~\ref{fig:MMS}{c}.
As MMS3 progresses across the separatrix, $\bar{f}(\vz,\vp)$ continues to change. At  times $\Delta t=226, 256$ ms,  the regions of incoming passing electrons with $\vz>0$ (labeled ${\bf B}$ in the
$\Delta t=16, 46, 76$ ms plots in Fig.~\ref{fig:MMS}{c}) are now dominated  by pitch angle mixed magnetosheath electrons streaming out along the separatrix from the reconnection region. These magnetosheath electrons are, naturally,  also subject to parallel acceleration (deceleration) by $\Phit$ and the mirror force, such that a fraction of these will be reflected back toward the diffusion region. Due to their larger density, the magnetosheath electrons dominate the full area in $(\vz,\vp)$-space labeled ${\bf D}$, previously occupied by trapped electrons and region ${\bf B}$ passing electrons.

%Particularly from $f(\vz,\vp)$ at $\Delta t=226$ ms in Fig.~\ref{fig:MMS}(c),  it is evident that  in the separator region, the magnetosheath electrons are mainly distributed on a shell of energies above a lower exclusion  energy $\cE_X$. Again, because of the pitch angle mixing in the diffusion region, the shell is uniform in the pitch angle $\theta\tan^{-1}(\vp/\vz)$, and is only interrupted by the trapped passing boundary marking the transition to the $\vz<0$ passing electrons originating in the magnetosphere.
%Below we derive a model for the cut-off energy, $U_c$, and thus, complete the description of this type of separator distributions characterized by a crescent in the $(\vz,\vp)$-plane.

The distributions in Fig.~\ref{fig:MMS}(d) are cuts through $f(\vz,v_{\perp1},v_{\perp2})$ with $\vz=0$. For $\Delta t= 76, 226$ ms  complete rings in the $(v_{\perp1},v_{\perp2})$-plane are clearly visible. Meanwhile, at $\Delta t= 46$ ms, the ring is incomplete and only a crescent is observed. For this location, the recorded magnetic field is relatively strong, $B=17.8$ nT, corresponding to a Larmor radius for the typical crescent electrons of less than 2km. %As will become clear below, these crescent electrons are  well magnetized  and are therefore not a direct measure of unmagnetized electron motion in the vicinity of the X-line.
As shown in Fig.~\ref{fig:MMS}(e), the distributions recorded by MMS4 are similar to those  of MMS3. However, the main and key difference (also discussed in Refs.~\cite{burch:2016,shay:2016}) is the reversed $\vz$-sign of the $(\vz,\vp)$-crescent for $\Delta t=136$ ms in Fig.~\ref{fig:MMS}(e). This is consistent with MMS4 crossing into the southern outflow, such that the $\vz<0$ passing electrons (region ${\bf C}$) are eliminated in favor of magnetosheath electrons streaming southward,   away from the diffusion region. %Again, these sheath electrons are subject to mirror and $\Phit$ deceleration. Conseqently, a significant fraction of them to reflect, such that the crescent continues beyond $\vz=0$ all the way to the boundary of the region ${\bf B}$ passing electrons.

\begin{figure}[h]
\centering
  \hspace{-0.5cm}\includegraphics[width=9cm]{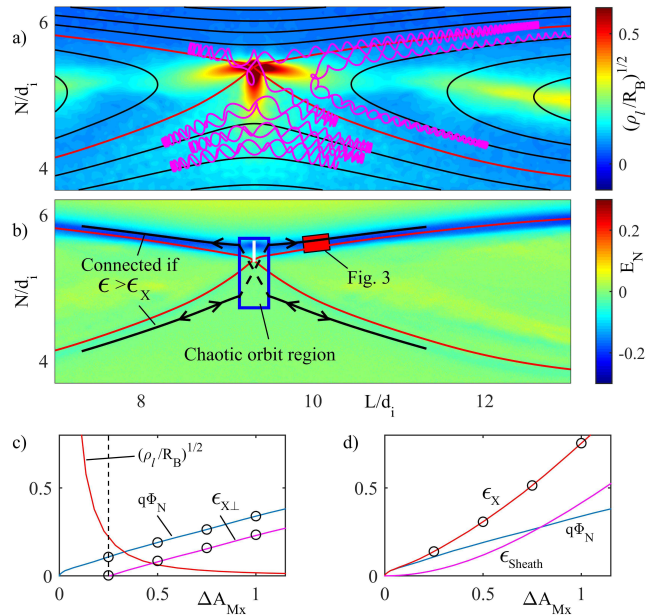}%
  \vspace{-0.4cm}
\caption{(a) Trapped magnetosheath electrons propagating across the diffusion region into the magnetospheric inflow. Non-gyrotropic regions ($\sqrt{\rho_l/R_B}>0.25$) are identified by the background color. (b) Guiding center trajectories  are characterized by $A_M=const$, and electrons with $\cE>\cE_X$ may jump  across the diffusion region. The color contours document the strong $E_N$ electric fields along the separatrix. (c) $q\Phi_N$ is a measure of the electron energization, obtained by integrating $E_N$ along the white line in (b). In areas of no pitch angle mixing (for $(\rho_l/R_B)^{1/2}<0.25$) the energization  $\cE_{X\perp}$ is purely perpendicular. (d) $\cE\scr{sheath}$ of Eq.~\eq{UL} and energization by $q\Phi_N$ add to provide the minimum kinetic energy $\cE_{X}$ of sheath electrons in the magnetopheric inflow. In (c,d), values are normalized by $mc^2$ and those used for computing the distributions in Fig.~\ref{fig:dists}(d) are marked by circles. }
\label{fig:doOrb}
\end{figure}

To explore the dynamics shaping the electron distribution function, we consider the trajectory in Fig.~\ref{fig:doOrb}(a) calculated using the magnetic and electric fields of a fully kinetic simulation (to be further described below). This electron enters the reconnection region on a trapped trajectory originating in the magnetosheath. It then travels into  the diffusion region and exits on the magnetopheric side of the separatrix. Only later does it reach the reconnection exhaust from the magnetospheric side.  Apart from the electron's brief encounters with the diffusion region, it is well magnetized with $\kappa^2=R_B/\rho_l\gg1$ \cite{buchner:1989}, where $R_B$ is the radius of curvature of the magnetic field and $\rho_l$ is the Larmor radius of the electrons.
On the other hand, the regions of chaotic unmagnetized electron dynamics  in Fig.~\ref{fig:doOrb}(a) are identified by the red areas where $1/\kappa=\sqrt{\rho_l/R_B}>0.25$. Here  the electron  magnetic moments $\mu=m\vp^2/(2B)$ are not conserved,  and their
pitch angles $\theta$ are randomized  \cite{le:2013,lavraud:2016}.

The above observations motivate  a model for the electron dynamics as sketched in Fig.~\ref{fig:doOrb}(b), where the electron motion outside the chaotic regions is described by the guiding center approximation. For 2D geometries, the canonical momentum in the out-of-plane $(M)$ direction of the guiding centers $P_{M,g}=qA_M + m\vz B_M/B$ is a constant of the guiding center motion. Here $A_M$ is the out-of-plane component of the magnetic vector potential, with the reconnection X-line characterized by the value $A_{M,x}$ observed at the saddle point in the profile of $A_M$.
Upstream and close to the separatrix, $B_M$ is small so that $P_{M,g}\simeq qA_M$, and it follows that guiding centers are locked to contours of constant $A_M$.

A quantitative condition required for magnetosheath electrons to jump to the  magnetospheric inflow region is obtained through the use of the canonical momentum $P_{M}=qA_M + mv_M$ of the full electron motion. This quantity is a constant of motion throughout the cross section,  including the chaotic orbit region. Variations in $mv_M$ determine the orbit size and  allow the Fig.~\ref{fig:doOrb}(a)   electron to move off  its particular  $A_M$-contour by the amount $\Delta A_M = m\Delta v_M/q$.
%In turn, because the  maximal value of $\Delta v_M$ is constrained by the electron kinetic energy, the maximal excursion $\Delta A_{M,max}$ obeys the equality $q^2(\Delta A_{M,max})^2/(2m)=\cE\scr{sheath}$.
Thus, a magnetosheath electron entering the chaotic region on a guiding center trajectory characterized by $P_{M,g}= qA_{M}$ can cross the separatrix only if the orbit permits a variation
\begin{equation}
\hblabel{eq:delA}
\Delta A_{Mx}\equiv A_M-A_{M,x}\quad.
\end{equation}
This requires an initial minimum kinetic energy in the magnetosheath given by
\begin{equation}
\hblabel{eq:UL}
\cE\scr{sheath}= \fd{q^2(\Delta A_{Mx})^2}{2m}\quad.
\end{equation}
Magnetosheath electrons passing through the diffusion region are energized by the strong $E_N$ electric field shown in Fig.~\ref{fig:doOrb}(b), characterized by $q\Phi_N$  in Fig.~\ref{fig:doOrb}(c), with $\Phi_N=-\int_0^{N(\Delta A_{Mx})} E_z dN$ evaluated along a cut starting at the X-line (short white line in Fig.~\ref{fig:doOrb}(b)).  The minimum  kinetic energy is then given by $\cE_X=q\Phi_N+\cE\scr{sheath}$, where   the $\cE\scr{sheath}$ contribution dominates for $\Delta A_{Mx}> 0.8$ (see Fig.~\ref{fig:doOrb}(d)). The energization by $E_N$ is in the perpendicular direction, but for $1/\kappa>0.25$ pitch angle mixing  transfers  a random fraction to the parallel direction for each electron. Meanwhile for $1/\kappa<0.25$, there is no pitch angle mixing such that the energization $\cE_{X\perp}$, identified in Fig.~\ref{fig:doOrb}(c), remains in the perpendicular direction. Thus,  any magnetosheath electron reaching sufficiently deep into the magnetospheric inflow will have a minimum perpendicular energy given by  $\cE_{X\perp}$, acquired outside the region of pitch angle mixing.

%The quantities $\cE_{X}$ and $\cE_{X\perp}$ thus  provide  minimum energy thresholds for  magnetosheath electrons observed in the magnetosphere. They are conveniently evaluated  as  functions of $\Delta A_{Mx}$  defined in Eq.~\eq{delA}, where $A_M$  is the value characterizing the guiding center trajectory on the magnetospheric side. The relevant value of $A_{M,x}(t)$ must be evaluated at the time the electron passes through the chaotic region.

%To emphasize the difference between the guiding center location and the actual electron location, we  consider a cut at fixed $L$. In this cut, let  $\Delta N_g$ represent the guiding center distance to the separator, such that $\Delta A_{Mx}\simeq q B_L\Delta N_g$. In terms of the electron location $\Delta N_g=\Delta N + m v_M/(qB_L)$ and it follows that $\Delta A_{Mx} = q B_L\Delta N + m v_M$. Spacecraft measure the electrons at their actual locations (at $\Delta N$), and, as we will discuss further below, the $m v_M$ correction (sign dependent on $v_M$) is fundamental for the occurrence of the perpendicular crescent distributions.

We may now derive a simple model for the drift kinetic \cite{kulsrud:1983} guiding center distribution $\bar{f}_{g}(\x_g,\vz,\vp)$ of magnetosheath electrons on the  magnetospheric side of the separatrix. Consistent with the simulation, we assume that the chaotic region is characterized by a Maxwellian distribution $f_{xline}(\cE)$. Using Liouville mapping of the phase density ($df/dt=0$),  it follows that
\begin{equation}
\hblabel{eq:fs}
\bar{f}_g\simeq  f_{xline}(\cE-q\Phi_N) \,   H(\cE-\cE_X)\,H(\cE_{\perp}-\cE_{X\perp})\quad,
\end{equation}
where $H(\cE)$ is the Heaviside step function. The heating by $q\Phi_N$ is included by evaluating $f_{xline}$ at $\cE-q\Phi_N$.
%As a likely explanation for the crescent shaped $(\vz,\vp)$-distributions observed by MMS, we note again that the sheath electrons streaming away from the chaotic region will fill all pitch angles except for those of the incoming passing electrons. %Also consistent with the MMS data, Eq.~\eq{UL} predicts that $\cE_X$ increases rapidly with the distance to the separator. %This is important for realizing sharp spatial gradients in $\bar{f}_s(\cE)$ required to explain the occurrence of crescent shapes in the $\vz=0$ cuts through the distributions.

\begin{SCfigure*}
%\begin{figure}[h]
\centering
  \includegraphics[width=12 cm]{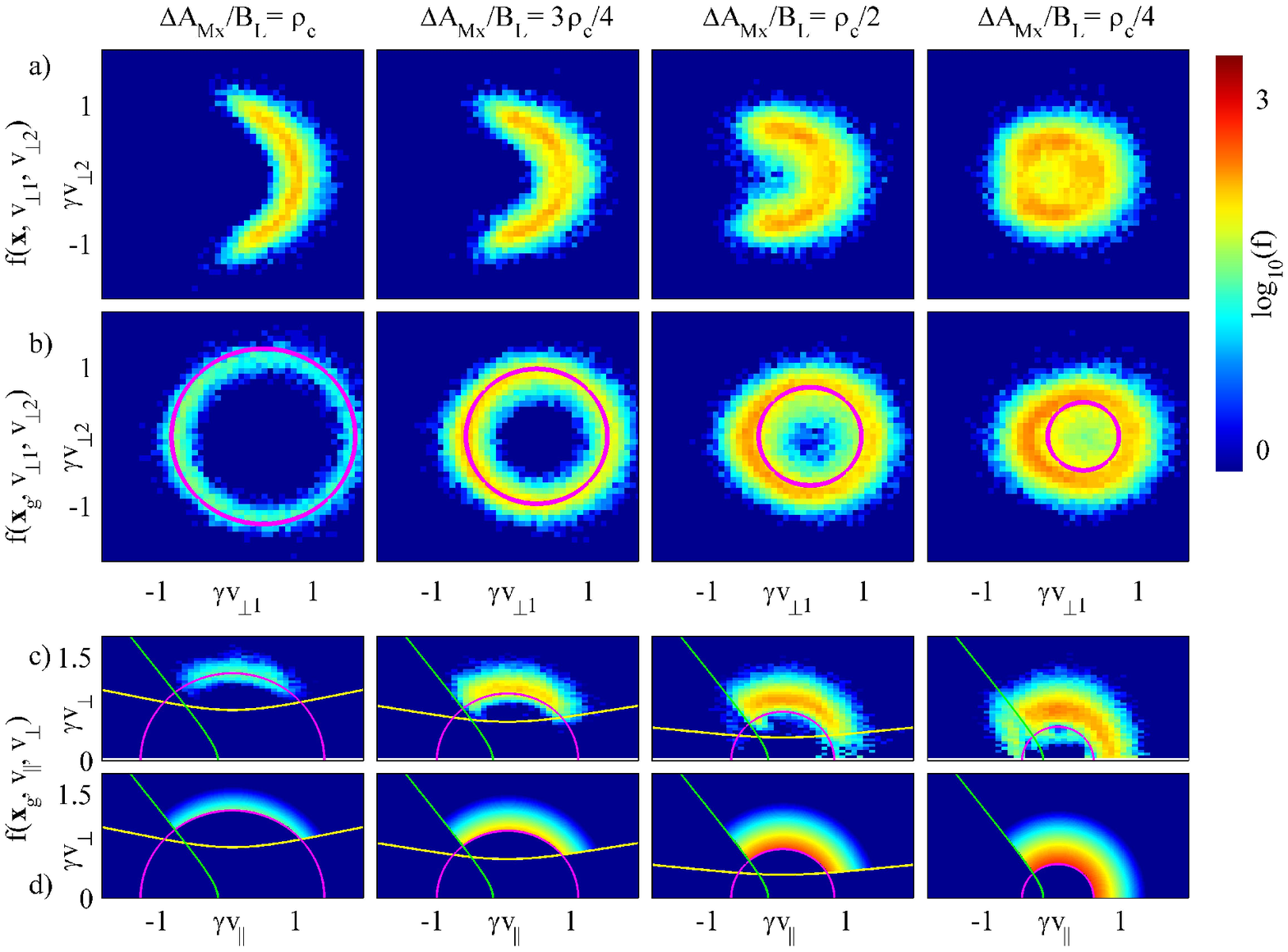}%
  \vspace{-0.6cm}
 \caption{(a,c) Rows of distributions for magnetosheath electrons collected from the location in Fig.~\ref{fig:doOrb}(b) as a function of $\Delta A_{Mx}$. Row (a) shows the local distribution $f(\x,v_{\perp1},v_{\perp2})= \int f(\x,\vv)\,d\vz$, while row (b) is calculated based on the location of the electrons' guiding centers ${f_g}(\x_g,v_{\perp1},v_{\perp2})= \int f_g(\x_g,\vv)\,d\vz$. In (b), the magenta lines show the predicted cutoff energy $\cE_X$ at values marked in Fig.~\ref{fig:doOrb}{d}, shifted by the E$\times$B-drift. Rows (c,d) display  $\bar{f}_g(\vz,\vp)$ obtained from the simulation data and Eq.~\eq{fs}, respectively. Again, the magenta lines represent the total cutoff energy $\cE_X$, while the yellow lines are the perpendicular cutoff energy $\cE_{X\perp}$; both calculated relativistically based on the values indicated in Figs.~\ref{fig:doOrb}{c}.
 The green lines mark the boundary to the velocity region of incoming passing magnetosphere electrons.
}
  %\vspace{-0.9cm}
\label{fig:dists}
%\end{figure}
\end{SCfigure*}

To validate the model in Eq.~\eq{fs}, we consider a kinetic simulation
performed with the VPIC code \cite{bowers:2009} using asymptotic plasma parameters identical to those of Ref.~\cite{burch:2016}. Here, however, the initial plasma current is carried by a modified Harris sheet \cite{roytershteyn:2012}. The reconnecting magnetic field and background temperatures  vary as $\tanh(N/d_i)$ ($d_i$ based on the magnetosheath density), and the density profile is adjusted to ensure magnetohydrodynamic pressure balance. The simulation is periodic in $L$ and has conducting boundaries in $N$, with a total size of $4032\times4032$ cells = $20\,d_i\times20\,d_i$. Separate populations of magnetosheath $(N>0)$ and magnetosphere $(N<0)$ particles with different numerical weights are loaded so that plasma mixing may be tracked over time \cite{daughton:2014} and so that both regions are resolved with ~400 particles per cell per species. Other simulation parameters are a mass ratio of $m_i/m_e=400$ and $\omega_{pe}/\omega_{ce}=1.5$ (based on magnetosheath field and density).

Particle distributions are computed at time $t=30/\omega_{ci}$, when reconnection has reached a quasi-steady state, and as indicated in Fig.~\ref{fig:doOrb}(b), we use electron  data collected
$0.7 d_i$ away from the X-line along the separatrix. The data includes only electrons originating from the magnetosheath side and is  collected as a function of $\Delta A_{Mx}$ for four locations within a narrow region reaching $\rho_c$ from the separatrix into the magnetospheric inflow. Here $\rho_c$ is the characteristic Larmor radius of a typical crescent electron (with relativistic momentum $m\gamma v\simeq mc$). Fig.~\ref{fig:dists}(a) shows the sequence of the full distributions integrated over the parallel velocity $f_{\perp}=\int f(\x,\vv) dv_{\parallel}$, revealing  crescent-shaped and ring distributions qualitatively consistent with the MMS observations. Meanwhile,  Fig.~\ref{fig:dists}(b) shows the sequence of distributions also integrated over the parallel velocity $f_{g\perp}(\x_g,\vv_{\perp})=\int f_g(\x_g,\vv)\, dv_{\parallel}$, but now with the numerical electrons binned  as a function of their guiding center locations. As such, $f_{g}(\x_g,\vv)$ is the distribution of guiding centers, defined without approximation through  $f_{g}(\x_g,\vv)\equiv f(\x_g-\boldsymbol\rho,\vv)$, where the direction of the vector $\boldsymbol\rho(\phi)=m\vv\times\B/(qB^2)$ is a function of the gyro-phase $\phi$.
The $f_{g\perp}$ distributions are characterized by nearly perfect circles in a frame slightly off-centered from the origin by the E$\times$B-drift; in this frame these crescent electrons follow nearly perfectly circular perpendicular gyro-orbits, and $f_{g\perp}(\x,\vv_{\perp})$ is nearly independent of $\phi$.

The gyro-averaged distribution of guiding centers  $\bar{f}_g(\x_g, \vz,\vp)$ in Fig.~\ref{fig:dists}(c) are  also compiled from the simulation particle data, where $\vp=(v_{\perp1}^2+v_{\perp2}^2)^{1/2}$ is evaluated in the frame of the E$\times$B-drift. %Thus, they obey  $\bar{f}_g(\x_g, \vz,\vp)=\left<f(\x_g-\bar{\rho}, \vv)\right>$, where $\left<\dots\right>$ denotes averaging over the gyro-phase $\phi$.
The exclusion energies $\cE_X$ of Fig.~\ref{fig:doOrb}(d) are shown by the magenta lines, accurately predicting the lower energy bound of the numerical distributions. The matching distributions in Fig.~\ref{fig:dists}(d) are obtained from Eq.~\eq{fs}, based on values of $q\Phi_N$, $\cE_{X\perp}$, and $\cE_{X}$ marked in Figs.~\ref{fig:doOrb}(c,d).  The combination of the exclusion energies  reproduces the behavior of the $(v_{\perp1},v_{\perp2})$-ring distributions with $\vz\simeq0$, evolving into crescent-shaped $\bar{f}_g( \vz,\vp)$-distributions for locations very close to the separatrix,  $\Delta A_{Mx}/B_L< \rho_c/2$.
We have verified that the model distribution in  Eq.~\eq{fs} is applicable along the separatrix of the full simulation domain, excluding only the region in Fig.~\ref{fig:doOrb}(b) where the electrons are unmagnetized. % (and $\bar{f}_g$ is not defined).

It is  evident from Fig.~\ref{fig:dists}(b-d) how $\cE_X$ rapidly increases with $\Delta A_{Mx}$, practically eliminating all electron guiding centers for   $\Delta A_{Mx} > B_L \rho_c$. However, the actual electron location is displaced from the guiding center  $\x = \x_g- \boldsymbol\rho$.  Depending on $\phi$, this allows electrons to penetrate up to an additional $\rho_c$ into the magnetospheric inflow. As the separatrix is approached from the magnetopause inflow,  the first magnetosheath electrons to be observed are those with $\phi$ placing their guiding centers closer to the separatrix.
As such, the crescent distributions are a manifestation of the diamagnetic drifts associated with the rapidly changing pressure of the magnetosheath electrons at the magnetopause/exhaust separatrix.

In summary, we have extended the analysis of the MMS electron data of Ref.~\cite{burch:2016} and shown that the observed  parallel heating of the magnetospheric inflow is consistent with the trapping model of  Refs.~\cite{egedal:jan2005,egedal:2008jgr}. Furthermore, the $(\vz,\vp)$-crescent distribution encountered by MMS can be accounted for by extending the electron dynamics of the trapping model to include magnetosheath electrons penetrating into the magnetophere. Here  the cutoff energy $\cE_X$
forbids electrons with insufficient diffusion region orbit size to reach into the magnetospheric inflow. The profile of $\cE_X$ depends strongly on the distance from  the separatrix, where the  chaotic region works like a {\sl Maxwell Demon}, only letting the most energetic magnetosheath electrons pass to the magnetospheric side. The perpendicular crescent-shaped distributions are formed due to the spatial gradients imposed by $\cE_X$.  They are a direct manifestation of the diamagnetic drift of well magnetized magnetosheath electrons in a boundary layer with a width of about an electron Larmor radius all along the low density separatrix.

%Because energetic electrons can  also pass in the opposite direction from the magnetopause to the magnetosheath, this is actually not a true Maxwell Demon, and the occurrence of the energetic ring distributions is of course not in breach of the  second law of thermodynamics

%\bibliographystyle{ieeetr}
%\bibliographystyle{apsrev4-1}
%\bibliographystyle{plainnat}
%\clearpage
%\bibliography{references}

\end{document}